# SPACE-TIME AS A DISCRETE FIELD NONCOMMUTATIVE CAUSAL NETWORK

**G.L. Stavraki**

Computing Centre of the Russian Academy of Sciences, Russia, 119333 Moscow, Vavilov Street 40.

**Abstract**

The necessity of rejecting the numerical model of geometrical extension is postulated on the basis of the idea of identity of space-time and physical vacuum. An attempt is made to define space-time not via the concept of manifold, but via the store of physical functions defined on it. The new description is based on the commutator representation of the causal structure of operator field theory. It is not the world point, but a universal field supermatrix complex $U$ that is assumed to be the carrier of possible local events. This complex involves a complete set of Heisenberg local field operators together with their spin-group bases in the Fermi-field representation. The fundamental element of the extension is described in the model by the equation of a special commutator algebra closed on two such local complexes $U_1$ and $U_2$ "nearest" in the two-sided light-like connection and linked by a single virtual field interaction vertex. The discrete character of the constructed "quantum proximity" equation containing the gravitational constant is associated with the existence of local curvature on the Planck scale. Algebraic closed-ness of the basic equation suggests that the charge symmetry group should be group $E_6$ with non-standard representations of the fermion and scalar fields. On the basis of the calculated U expression we propose an effective superinvariant Lagrangian with fixed coefficients on the near-Planck scale, from which one can in principle try to obtain a low-energy limit for comparison with the real world.

**1. Introduction. Physical scheme**

The idea of geometrization of some elementary particle physics notions on the basis of the concept of manifold, which has been fruitful over many years, seems to reveal arbitrariness in the corresponding (string-membrane) constructions on very small scales. On the other hand, it must have obviously been commonly agreed that at such distances the idea of the geometrical structure of the world must somehow be changed. That is why, in this work we develop, in line with [1], the approach based on the representation of the primary space-time as a set of virtual local events with a unique a priory structure, namely, causal ordering.

As a primary way of describing such a World beyond the Lagrangian formalism the Heisenberg picture is used in the framework of the axiomatic Wightman approach in which the interacting quantum fields are described by local operators carrying all the information of possible events in the neighborhood of each point [2]. Such fields are universal, indestructible, and cannot essentially be separated from space-time. In particular, creation and annihilation of the corresponding quanta seem to be the "handwork" of space-time itself rather than something additionally embedded in it. In our construction we shall therefore proceed from the attempt to identify space-time and what is called the physical vacuum. Within this approach, such Heisenberg field operators are taken together with their spin-group bases in the Fermi-field representation and they should be regarded as primary geometrical elements of a certain new – "physical" – geometry uniting both geometry and physics. The World 4-points are replaced in this physic-geometrical scheme by complete (presumably finite) specially organized local sets of field linear forms further on referred to as $U$ complexes. A detailed description of the structures of these complexes is given below.

Within such a general approach, as a substitute for concrete equations of motion we adopt here the conception of the causal structure of space-time described in the framework of the classical picture in books [3], [4], [5] and proceeding [6]. In this sense it is essentially important that on the set of Heisenberg field operators such a structure in quantum field theory is universally defined by the relations $[A(1), B(2)]_{\pm} = 0$ for fields *A* and *B* at causally disconnected (mutually spacelike) points 1~2 and $[A(1), B(2)]_{\pm} \neq 0$ for interacting fields *A* and *B* at causally connected points, where $[..,..]_{\pm}$ is a physically graded (anti)commutator ($[,]_{+}$ for arguments-fermions and $[,]_{-}$ otherwise) [2].

On the other hand, in the local instantaneously static situation with zero intrinsic time typical of massless fields, a virtual act of local field interaction is taken as the only type of possible events.

The next step in the model is the assumption that the whole set of quantum fields forms a virtual causal "medium" in the sense that the motion-development is only possible here as a "short-range action"), i.e., exclusively through acts of field interactions (rather than motions) of two causally nearest complexes. The reason for it is assumed to be a virtual self-closure of space-time on Planck's scale in the neighborhood of each local interaction of primary physical fields as a result of an accompanying unlimited increase of energy density fluctuations (and gauge charge!) at a "given place". Each pair $U_1 \prec U_2$, where $\prec$ is causal ordering and the value of time 0-component of the non-spacelike segment $(1,2)$ is of the order of $\leq \sqrt{g}$ ($g$ is the gravitational constant), is assigned one and only one field interaction act which, from the viewpoint of the classical geometrical picture, forms the vertex of a two-cavity light cone with compactified cavities, $U_1$ being put into correspondence with the cavity of the "nearest" local past and $U_2$ - with the "nearest" local future in the sense specified in the course of construction. Furthermore, such a unified deformation of the world neighborhood of each act is assumed in the model to be the only primary manifestation of gravity which is understood as the reaction of space-time to the virtual field interactions proceeding in it and which in a submicroworld is therefore not a field, as distinct from the fields entering in the $U$-complex, but rather a universal local geometric effect of local curvature of the causal structure accompanying any local change in the virtual physical world. From this point of view, large-scale gravity must occur in the model as an effective connectedness field on the set of local frame of reference to which the primary physical fields, i.e., components of the supermatrix $U$, refer. The whole space-time is regarded as a system of elementary compactified cells-cones represented as a certain algebraic structure connecting two causally nearest (by definition) field $U$-complexes and containing Lorentz-anisotropic metric factors which presumably correspond to the gravitational-inertial forces generated by the two-point closure of the algebraic equation. The very existence of separate elements of the structure of virtual World can then be considered as a result of occurrence in each interaction act of local "instantaneous" D'Alembert-Poincare equilibrium between the field (charge) and gravitational-inertial forces, which is called on, in particular, to provide equality to zero of the intrinsic tension of field quanta.

This work is an attempt to formulate the description of such a presumed fundamental causal cell of proximity and local equilibrium on the basis of correspondence principle with Lagrangian operator field theory.

From the viewpoint of the presented approach, what is to be done to make a concrete model of it is to replace the world points by appropriate local field complexes, to define fields in them as corresponding averaged operators of the Lagrangian scheme, and using the causality principle in the form

$$[A(1), B(2)]_\pm \neq 0,$$

where $A$ and $B$ are interacting fields, the segment $(1,2)$ is non-spacelike, and $[..,..]_\pm$ is a physically graded (anti)commutator, to construct, in fact, an algebraic noncommutative equation which would in a sense correspond to very small scales and define on the set of such complexes the binary ratio of absolute (maximum) causal proximity.

## 2. Description of the structure of the local field complex $U$.

As can be seen from the general description, the model is based on the representation of space-time by the store of functions-fields existent in it. That is why the model has no coordinates and coordinate "background" bases. Instead, finite-dimensional group bases of field are used in the representation of Fermi fields which by definition interact with all the other fields and in this sense are assumed to be fundamental. Accordingly, the $U$-complex involves the fields convolute with their matrix bases. On the other hand, the understanding some time ago of unified algebraic nature of Bose and Fermi fields in [7], [8], and [9], which underlay the conception of supersymmetry ([10]), allows considering them within the model as different components of a specially organized supermatrix $U$ [11] that realize the representations of the Lorentz spin group $L_{sp}$ and the charge symmetry group $\Omega$:

$$U = \begin{pmatrix} \left(\kappa_i \cdot \delta_{ij}\right) \times \underset{\downarrow}{\begin{pmatrix} u_{11} & u_{12} \\ u_{21} & u_{22} \end{pmatrix}}^{\rightarrow} & \kappa_i \cdot \underset{\downarrow}{\begin{pmatrix} u^i_{13} \\ u^i_{23} \end{pmatrix}}^{\Rightarrow} \\ \kappa_j \cdot \underset{\Downarrow}{\left(u^j_{31}, u^j_{32}\right)}^{\rightarrow} & \underset{\Downarrow}{(0)}^{\Rightarrow} \end{pmatrix} \quad \{1\}$$

The block $\begin{pmatrix} u_{11} & u_{12} \\ u_{21} & u_{22} \end{pmatrix}$ is composed of Bose fields, the blocks $\begin{pmatrix} u^i_{13} \\ u^i_{23} \end{pmatrix}, \left(u^j_{31} \ u^j_{32}\right), i, j = I, II, III$ consist of Fermi fields, the superscripts $i, j$ run through identical fermion families noninteracting among themselves, and $\kappa_I, \kappa_{II}, \kappa_{III}$ are three different numerical roots of the equation $\kappa^3 = 1$ ( $k_I + k_{II} + k_{III} = 0$, $k_I^2 + k_{II}^2 + k_{III}^2 = 0$, $k_i^2 = \widetilde{k}_i$, and $\widetilde{(\ )}$ is complex conjugation), and the block $\left(k_i \cdot \delta_{ij}\right)$ is the $i, j$-matrix. We shall see below that it is precisely such an introduction of precisely three fermion families that provides closure of the proximity algebra introduced later in the framework of the correspondence principle with the Lagrangian formalism.

Spin 0,1 and $\frac{1}{2}$ fields are represented in the $U$-matrix by the matrix forms linear in them in the representation of Fermi fields of the group $L_{sp} \times \Omega$. The arrows $\downarrow, \rightarrow$ designate running combined spin-charge indices of fermion representation in this group and the arrows $\Downarrow, \Rightarrow$ stand for running Lorentzian vector indices.

The explicit form of the blocks-submatrices (u) is

$$u_{11} = a_{11}\left(\frac{I+\gamma^5}{2}\sigma^{\lambda\mu}t_s\right)F_{\lambda\mu,s} + \frac{I+\gamma^5}{2}\left(y_1\delta_{r_1 r_2}I + y_2 q_{r_1}q_{r_2}^+\right)\varphi_{r_1}\varphi_{r_2}^+, \quad u_{12} = -ia_{12}\left(\frac{I+\gamma^5}{2}\gamma^\lambda C q_r\right)\varphi_{\lambda,r}$$

$$u_{22} = a_{22}\left(\frac{I-\gamma^{5T}}{2}\sigma^{\lambda\mu T}t_s^T\right)F_{\lambda\mu,s} + \frac{I-\gamma^{5T}}{2}\left(z_1\delta_{r_1 r_2}I + z_2 q_{r_2}^+ q_{r_1}\right)\varphi_{r_1}\varphi_{r_2}^+, \quad u_{21} = -ia_{21}\left(\frac{I-\gamma^{5T}}{2}C^{-1}\gamma^\lambda q_r^+\right)\varphi_{\lambda,r}^+ \quad \{2\}$$

$u_{13} = a_{13}(\gamma^\lambda \psi), \; u_{23} = a_{23}(\bar{\psi}\gamma^\lambda), \; u_{31} = -ia_{31}\bar{\psi}_\lambda + c_{31}C^{-1}\gamma_\lambda q_r^+ \psi \varphi_r^+, \; u_{32} = -ia_{32}\psi_\lambda + c_{32}\gamma_\lambda C q_r \bar{\psi}\varphi_r$

Within the Lagrangian formalism, the corresponding field variables are interpreted as follows.

$F_{\mu\nu,s}$ are tensions of the $s$-multiplet of gauge fields $A_{\mu,s}$,

$$F_{\mu\nu,s} = \partial_\mu A_{\nu,s} - \partial_\nu A_{\mu,s} - E_{ss_1 s_2} A_{\mu,s} A_{\nu,s}$$

$\varphi_{\lambda,r} = D_\lambda \varphi_r = \left(\partial_\mu - iT^\varphi_{s,r_1 r} A_{\mu,s}\right)\varphi_{r_1}, \; \varphi_{\lambda,r}^+$ are gauge derivatives of $r$-multiplet of scalar fields $\varphi_r, \varphi_r^+$,

$\psi_{\lambda,n} = \left(\partial_\lambda - it_{s,nn_1}A_{\lambda,s}\right)\psi_{n_1}, \; \bar{\psi}_{\lambda,n}$ are gauge derivatives of $n$-multiplet of any single family of left spinor $(\lambda)$ $\frac{1}{2}$-spin fermion fields,

$$\psi_n \equiv \frac{I-\gamma^5}{2}\psi_n, \quad \bar{\psi}_n \equiv \bar{\psi}_n \frac{I+\gamma^5}{2}.$$

$X_s = t_s, T^\varphi_{s,r r_1}$ are Hermitian generators of the Lie algebra $X_{s_1}X_{s_2} - X_{s_2}X_{s_1} = -iE_{s_1 s_2 s_3}X_{s_3}$ of the charge symmetry group $\Omega$ in respectively the fermion and scalar field representations, and $E_{s_1 s_2 s_3}$ are real structural constants totally antisymmetric in the subscripts $(s_l)$. The matrices $q_r$ satisfy the group covariance relation $t_s q_r + q_r t_s^T = T_{s,r_1 r}^\varphi q_{r_1}$. The group matrices are assumed to be normalized by the relations

$$E_{ss_1s_2} E_{ss_1s_3} = b_1 \delta_{s_2 s_3}, \quad (T_s^\varphi T_s^\varphi)_{rr_1} = b_3 \delta_{rr_1}, \quad (t_s t_s)_{nn_1} = b_4 \delta_{nn_1}, \quad (q_r q_r^+)_{nn_1} = b_5 \delta_{nn_1}, \quad b_{1,3,4,5} > 0 \quad \{3\}$$

and to satisfy additionally the equation $T_{s,rr_1}^\varphi q_r q_{r_1}^+ = b_2 t_s$, $b_2 = \dfrac{b_3 b_5}{2 b_4}$, which turns out to be one of the necessary conditions of the existence of the corresponding proximity equations. The numerical coefficients $a_{..}, c_{..}, y, z$ in the initial expression for the $U$-matrix are unknown. Their determination, along with the group parameters $b_{1...5}$, from the requirement that the presumptive proximity equation should be satisfied by this expression within the framework of the correspondence principle with Lagrangian formalism of quantum field theory was the primary intent of the work. In the calculation we used the field Lagrangian density in the first-order formalism:

$$L = \frac{i}{2}\left(\bar{\psi}^{(i)} \gamma^\mu D_\mu \psi^{(i)} - D_\mu \bar{\psi}^{(i)} \gamma^\mu \psi^{(i)}\right) - \frac{1}{2}\left(\bar{\psi}^{(i)} q_r C \bar{\psi}^{(i)} \varphi_r - \psi^{(i)} q_r^+ C^{-1} \psi^{(i)} \varphi_r^+\right) + D_\nu \varphi_r \overset{+}{\varphi}_r^\nu + D_\nu \overset{+}{\varphi}_r \varphi_r^\nu - \overset{+}{\varphi}_r^\mu \varphi_{\mu,r} + X_{r_1 r_2 r_3 r_4} \varphi_{r_1} \varphi_{r_2} \overset{+}{\varphi}_{r_3} \overset{+}{\varphi}_{r_4} + \frac{1}{4} F_{\mu\nu,s} F_s^{\mu\nu} - \frac{1}{2} F_s^{\mu\nu}\left(\partial_\mu A_{\nu,s} - \partial_\nu A_{\mu,s} - E_{ss_1 s_2} A_{\mu,s_1} A_{\nu,s_2}\right), \quad \{4\}$$

where $D_\mu$ are gauge derivatives, $C$ is the spinor matrix of charge conjugation, and $X_{...}$ is the $\Omega$-covariant vertex matrix of scalar-field self-action.

## 3. Assumed light-likeness of primary causal relationships at supersmall distances. Light-like canonical quantization.

In the model construction on the basis of the classical Lagrangian geometrical picture we used the essential assumption concerning light-likeness of a supersmall neighborhood of a local virtual field interaction act. The world points connected by it are assumed to belong to one of the light-cone generators. Hence, in the Lagrangian formalism of the correspondence principle canonical quantization on the light cone should be used ([12], [13], [14]). In its framework, the Minkowski space $\left(g_{\mu\mu} = g^{\mu\mu} = (1,-1,-1,-1)\right)$ is represented by a sheaf of parallel light-like hyperplanes $n_{+,\nu} x^{\nu} = const$, where $n_{+,\mu}$ is the 4-vector from the pair of light-like vectors $n_{\pm,\mu}$ with the properties $n_{\pm}^{\mu} n_{\pm,\mu} = 0$, $n_{\pm}^{\mu} n_{\mp,\mu} = 1$, and $n_{\pm}^{\mu} = n_{\mp,\mu}$. An arbitrary pair of free light vectors $n'_{\mu} \neq n''_{\mu}$, $n'_{\mu} n'^{\mu} = 0$, $n''_{\mu} n''^{\mu} = 0$, with the property $n'_{\mu} n''^{\mu} > 0$ is reduced, by the choice of Lorentzian reference frame with the time axis along the vector $n'_{\mu} + n''_{\mu}$ and an appropriate normalization, to the form of the pair $n_{+,\mu}$, $n_{-,\mu}$ which in this sense is the case of general position.

Via the identical decomposition

$$g^{\mu\nu} \equiv g^{\perp\mu,\perp\nu} + n_{+}^{\mu} n_{-}^{\nu} + n_{-}^{\mu} n_{+}^{\nu}, \quad x^{\mu} \equiv x^{\perp\mu} + n_{+}^{\mu} \cdot u + n_{-}^{\mu} \cdot V, \quad u = n_{-,\nu} x^{\nu}, \quad V = n_{+,\nu} x^{\nu} \quad \{5\}$$

$x^{\mu} = \left(x^{(0)}, x^{(1)}, x^{(2)}, x^{(3)}\right)$ are replaced by the so-called light coordinate variables $\left(V, u, x^{\perp\mu}\right)$

(in a particular realization

$$V = \frac{1}{\sqrt{2}}\left(x^{(0)} + x^{(3)}\right), \quad u = \frac{1}{\sqrt{2}}\left(x^{(0)} - x^{(3)}\right), \quad x^{\perp\mu} = \left(0, x^{(1)}, x^{(2)}, 0\right) \quad ). \qquad \{6\}$$

$V$ plays the role of time variable, and the canonical (anti)commutators of fields $A$ on the light-like hypersurfaces $V = const$ are constructed as

$$\left[A_{\alpha}(1), p_{(A_{\beta})}^{(V)}(2)\right]_{\pm}\bigg|_{V_1 = V_2} = \frac{i}{2}(I_A)_{\alpha\beta} \delta(u_1 - u_2) \delta^{(2)}\left(x_1^{\perp\lambda} - x_2^{\perp\lambda}\right)$$

where $p_{(A_{\alpha})}^{(V)} = \frac{\partial L}{\partial(\partial_V A_{\alpha})}$ and $I_A$ is the matrix projection operator

$I_A A \equiv A$, $p_A^{(V)} I_A \equiv p_A^{(V)}$, $I_A^2 \equiv I_A$ corresponding to the multicomponent field $A$.

Such a representation, in particular, of a geometrical space will in a sense be considered as fundamental and, accordingly, in the Lagrangian we shall pass over to the description of fields in light variables. We shall also assume that an adequate representation in it of a gauge field as a dynamical system must be analogous to the other fields - without relation in the operator part of the description and with the zero corresponding Lagrange factor as an extra nonphysical variable. Assuming also that along the light $u$-generator of relativistic cone of causality the phase will not change, i.e., $\int_{(u)} A_{\mu,s} dx^\mu = 0$, we obtain the gauge condition on the vector fields in the form $n_{+,\mu} A_s^\mu = 0$ or $A_s^\mu \equiv \left(g_\nu^\mu - n_-^\mu n_{+,\nu}\right) A_s^\nu$.

The canonical "coordinate" variables in the description of vector fields will be $A_{\perp\mu,s} = \left(g_\mu^\nu - n_{+,\mu} n_-^\nu\right) A_{\nu,s}$, the "momentum" variables will be $F^{\perp\mu}{}_{,u,s} \equiv \left(g_\nu^\mu - n_+^\mu n_{-,\nu}\right) \cdot n_{+,\lambda} F^{\nu\lambda} = -\partial_u A_s^{\perp\mu}$, and $I_A = g_{\perp\mu}^{\perp\nu} \cdot \delta_{ss_1}$. For scalar fields the canonical pairs will correspondingly be $\varphi_r$ and $\varphi_{u,r}^+$ ($I_\varphi = \delta_{rr_1}$) and for spinor fields - $\psi_V = p_V \psi$ and $\frac{i}{2}\overline{\psi}_V \gamma_V$, where $p_V = \frac{1}{2} n_{-,\mu} n_{+,\nu} \gamma^\mu \gamma^\nu$, $\gamma_V = n_{+,\mu} \gamma^\mu$, and $I_\psi = \frac{I-\gamma^5}{2} p_V \cdot \delta_{nn_1}$. All the other field components are dependent and are determined from equations of motion of the form $\partial_u Z = R$ as $Z = (\partial_u)^{-1} R$. In line with [14] we choose the inverse integral operator $(\partial_u)^{-1}$ with a kernel antisymmetric in arguments, as in $\partial_u$ itself, and take it as a universal representation of the dependent field components of the correspondence principle in this model. As a result, we obtain

$$Z = \frac{1}{2} \int_{-\infty}^{\infty} \varepsilon(u-\xi) R(\xi) d\xi, \qquad \varepsilon(u-\xi) = \begin{cases} 1 & u > \xi \\ -1 & u < \xi \end{cases} \qquad \{7\}$$

$\partial_u \varepsilon(u-\xi) = 2\delta(u-\xi)$, with the boundary condition
$Z\left(u = -\infty, V, x^{\perp\mu}\right) = -Z\left(u = \infty, V, x^{\perp\mu}\right)$.

The distribution value of such field components along (u) is determined in the model by their established interaction-connection with the distribution of the corresponding sources for which these fields are proper.

The same rule is applied to restore the commutators $\left[\varphi_r(1),\varphi^+_{r_1}(2)\right]_-$ from $\left[\varphi_r(1),\partial_u\varphi^+_{r_1}(2)\right]_-$ and $\left[A_{\perp\mu,s}(1),A_{\perp\nu,s_1}(2)\right]_-$ from $\left[A_{\perp\mu,s}(1),\partial_u A^{\perp\nu}_{s_1}(2)\right]_-$ in the canonical commutation relations.

As a result of construction of the correspondence principle, such antisymmetry of the integral kernel leads in the assumed fundamental algebra to antisymmetry of the model bilinear algebraic operation in its arguments – field complexes at two points – the ends of an elementary link of the operator network.

The constructed formalism with $V$ as "time" and $u$ as an analogue of the spatial variable (along with $x^{\perp\mu}$) will be called the $L_u$-formalism.

In view of the perfect symmetry $V \leftrightarrow u$ of the variables, on the other family of parallel light-like hyperplanes: $u' = n_{-,\nu}x^\nu = const$ corresponding to the "time" $u'$ and the "spatial" variable $V'$, one can construct an $L_V$-formalism, dual to $L_u$, with phase conservation along the $V'$- generator, the corresponding gauge condition $n_{-,\nu}A'^\nu_s = 0$ $\left(A'^\mu_s \equiv \left(g^\mu_\nu - n^\mu_+ n_{-,\nu}\right)A'^\nu_s\right)$, and the integral representation of the dependent field components:

$$Z' = \left(\partial^{-1}_{V'}\right)R' = \frac{1}{2}\int_{-\infty}^{\infty}\varepsilon(V'-\xi)R'd\xi \qquad \{8\}$$

Both these canonical formalisms, together with the corresponding 4-space divisions are conjugated with each other relative to the hyperplane $\left(x^{(0)}=0,x^{(1)},x^{(2)},x^{(3)}\right)$ via $T$-reflection whose constituent is an antiunitary involutive time reflection operator on the quantum fields of Lagrangian $\{4\}$ ([15], [16]).

On the one hand, this $T$-transformation relates geometrically the points of $u$- and $V$-generators of cone and, on the other hand – the fields of $L_u$- and $L_V$-formalisms at these points.

**4. Conclusions from the assumed physical picture, necessary for formulation of the correspondence principle of the model with**

**quantum field theory. *T*-symmetry of the fundamental algebra. Averaged field operators. Definition and approximate calculation of bilinear algebraic operation on field complexes on two *T*-conjugate light generators and closure – in fields – of the corresponding separate $L_u$ and $L_v$- algebras. Four-point *T*-invariant superposition of $L_u$ and $L_v$ algebraic structures.**

In the employed conditional physical picture of submicroscopic structure, each discretely separated elementary link in the unified causal network corresponds to an ultrashort existence of a local elementary object – a field complex – within a black hole which, as has already been said, is presumably due to an unlimitedly increasing energy-density and gauge-charge quantum fluctuations at the moment and place of a local virtual interaction act. In line with this standpoint, a physical construction of neighborhood such an act is taken as a set of absolute (light-like) motions directed towards or from the source (interaction point). As a result of physical separation of event flow in local reference frames falling towards the black hole center in different directions near its horizon one can think that fields at a submicroscopic level can be referred to the reference frame ("instantaneously" proper?) in which the source radiation is "needle"-shaped, i.e., a given virtual interaction act is connected with one and only one *T*-symmetric pair of light-like segments intersecting in it. Since each elementary object within such a mini-hole is physically isolated, its dynamical state which is due only to a given field interaction act can be represented as corresponding to local force equilibrium and, accordingly, as instantaneously static or, in other words, *T*-symmetric relative to some local time. In the described picture such a state must obviously be described only through superposition of two light-like motions, namely, towards the source and away from it. The conclusion is that the corresponding supposed model proximity algebra should be based on the description of the two light-like motions characterizing the dual nature of the source (sink ↔ source proper) and at the same time it should be $T$-symmetric. Some analogy to this may be the well-known situation with the Dirac equation in the Hamiltonian form in which the eigenvalues of the matrix velocity operator will be $\pm c$ and the state corresponding to any velocity $|v|<c$ appears only as a result of superposition of eigenstates.

One of the light-like motions in the model is associated with the $u$-generator and $L_u$-field formalism and the other motion, opposite to the first one – with the $V$-generator and $L_V$-field formalism.

According to the above arguments within the framework of the correspondence principle an isolated elementary object – a field complex – must locally be completely concentrated on the $u,V$-generators as light trajectories. Hence, in the algebra construction we used complexes of the form $\{1\}$, $\{2\}$ which are conditionally averaged over a small $(\sim g)$ finite region $x^{\perp\mu}$ of 2-plane. If $\breve{U}$ is an averaged complex and $\sigma$ is the domain of $x^{\perp\mu}$ variation restricted by the condition

$$\int_\sigma d^{(2)}x^\perp \simeq g, \quad \text{then} \quad \breve{U} = \frac{1}{g}\int_\sigma U \cdot d^{(2)}x^\perp. \qquad \{9\}$$

We also assume $\int_\sigma \partial_{\perp\mu} A \cdot d^{(2)}x^\perp \simeq 0$, where $A$ is an arbitrary operator and $\int_\sigma \delta^{(2)}\left(x^{\perp\lambda} - y^{\perp\lambda}\right) \cdot d^{(2)}x^\perp = 1$.

The algebraic relation between the causal shift and field interactions is in principle well known and consists in the fact that the result of the physically graded commutation $\left[A(1),B(2)\right]_\pm$ of the Heisenberg fields $A(1)$ and $B(2)$ entering the vertex $\sim ABC$ at causally connected points $(1)\prec(2)$ contains the operator contribution due to the field $C$. In some cases, in the first approximation with respect to the causal shift this contribution can even be represented directly as $\left[A(1),B(2)\right]_\pm \simeq \sim I + \sim \left(C(1)+C(2)\right)$. This circumstance suggests that field interaction beyond the framework of Lagrangian formalism can be described purely algebraically in the spirit of the principle "two (causally separated) fields from 3-vertex generate a third field": $AB \to C$, $BC \to A$, $AC \to B$. One can see that such a description corresponds schematically to the structure of a bilinear closed algebra (some time ago this led to the idea of Lie superalgebras [8]). The main assumption following from this is the possibility of field-algebraic description of causal proximity.

According to this assumption, within the framework of the $L_u$-formalism we shall define a bilinear antisymmetric operation on the complexes $\breve{U}_1, \breve{U}_2$, corresponding to two different $u$- points $(1),(2) \in (u)$ $(1) \prec (0) \prec (2)$ of the generator $V = 0$ ( $(0)$ is the intersection point of $u$ and $V$-generators) in the form [17]

$$\{\breve{U}_1, \breve{U}_2\}\big|_{V=0} \equiv -4ig\left( \left[\breve{U}_1\big|_{V=0}, \breve{U}_2\big|_{V=0}\right]_- + \left[\breve{U}_1^\tau\big|_{V=0}, \breve{U}_2^\tau\big|_{V=0}\right]_-^{(\tau^{-1})} \right) \quad \{10\}$$

Here $g$ is the constant of dimensionality $[l]^2$, which we identify with the gravitational constant and $(\ )^\tau$, $(\ )^{(\tau^{-1})}$ are supertranspositions ([11]) of cell supermatrix $\begin{pmatrix} B_1 & F_1 \\ F_2 & B_2 \end{pmatrix}$ of the form $\begin{pmatrix} B_1 & F_1 \\ F_2 & B_2 \end{pmatrix}^\tau \equiv \begin{pmatrix} B_1^T & F_2^T \\ -F_1^T & B_2^T \end{pmatrix}$, $\begin{pmatrix} B_1 & F_1 \\ F_2 & B_2 \end{pmatrix}^{(\tau^{-1})} \equiv \begin{pmatrix} B_1^T & -F_2^T \\ F_1^T & B_2^T \end{pmatrix}$, where $(\ )^T$ is an ordinary transposition involving only the matrix bases of cells but not the field operators ($B_i = \sum_\alpha \Gamma_{i,\alpha}^b b_{i,\alpha}$, $F_i = \sum_\alpha \Gamma_{i,\alpha}^f f_{i,\alpha}$ are matrix forms with boson and fermion fields $b_{i,\alpha}, f_{i,\alpha}$).

Owing to the special structure of the operation $\{10\}$ it can identically be written ([17]) as a supermatrix composed of field expressions of the type

$$\left(\left[\breve{b}_{i,\alpha}(1), \breve{b}_{j,\beta}(2)\right]_- - (1) \leftrightarrow (2)\right)\bigg|_{V=0}, \quad \left(\left[\breve{b}_{i,\alpha}(1), \breve{f}_{j,\beta}(2)\right]_- - (1) \leftrightarrow (2)\right)\bigg|_{V=0},$$

$$\left(\left[\breve{f}_{i,\alpha}(1), \breve{f}_{j,\beta}(2)\right]_+ - (1) \leftrightarrow (2)\right)\bigg|_{V=0}.$$

In the lowest (zero!) approximation in $|(2)-(1)| \equiv u_2 - u_1 > 0$, with a restriction to single-vertex contributions and with allowance for conditions {9} these expressions antisymmetric with respect to the permutation $(1) \leftrightarrow (2)$ appear (with one exception) to be proportional to the expressions

$$\sim \varepsilon(1,2)\big(\breve{b}(1)+\breve{b}(2)\big), \quad \sim \varepsilon(1,2)\big(\breve{f}(1)+\breve{f}(2)\big), \ (\ \varepsilon(1,2) \equiv \varepsilon(u_1 - u_2) = -1 \text{ under condition } (1) \prec (2)\ ) \quad \{11\}$$

i.e., are expressed in terms of the same fields as the arguments of operation {10}. The exception from such algebraic field closure concerns the cell $\big(\{\ \}\big|_{V=0}\big)_{33}$, which leads, in particular, to gauge-noncovariant contributions $\sim \big(A_{\perp\mu,s} A_{\perp\nu,s}(1) + A_{\perp\mu,s} A_{\perp\nu,s}(2)\big)$. This undesirable contribution is eliminated as a whole through the introduction (performed in a special way and described in Sec. 2) into the U structure of exactly three identical fermion families noninteracting among themselves directly. As a result we arrive at $\big(\{\ \}\big|_{V=0}\big)_{33} = 0$, and the algebra with the operation $\{\ \}\big|_{V=0}$ in the indicated approximation appears to be closed with respect to the fields entering in the $U$-composition {2} taken at points $(1)$ and $(2)$.

The expression $\{\breve{U}_{1'},\breve{U}_{2'}\}\big|_{u=0}$, $T$-dual to {10}, with analogous results of approximate calculation is determined within the framework of $L_V$-formalism for points $(1'),(2') \in (V)$, $(1') \prec (0) \prec (2')$ of the $V$-generator of cone.

It now seems natural to determine the $T$-invariant superposition of contributions into the operation of presumed algebra from the two light-like motions along the intersecting u- and V-generators:

$$\frac{1}{2}\Big(\{\breve{U}_1,\breve{U}_2\}\big|_{V=0} + \{\breve{U}_{1'},\breve{U}_{2'}\}\big|_{u=0}\Big) \quad \{12\}$$

where points $(1)$ and $(1')$, $(2)$ and $(2')$ are mutually spacelike ($(1) \sim (1'), (2) \sim (2')$) and points $(1)$ and $(2'), (2)$ and $(1')$ are $T$-conjugated between themselves relative to the $T$-invariant hyperplane $(n_{+,\mu} + n_{-,\mu})x^\mu = 0$. According to the above arguments, the result of the approximate calculation of this expression is expressed by a linear matrix combination of fields-arguments defined at these four points.

**5. $U$-complex as a primary local object. Requirement of closure of the fundamental algebra in terms of $U$-complexes and its realization through the narrowing of equation {12} onto two points. Interpretation of the corresponding identification of fields as a jump-like transition to a locally curved world consistent with the GR equivalence principle. Closure of a two-point algebra as a non-Lagrangian inclusion of field interactions. Explicit form of the abstract fundamental algebra and its possible physical meaning.**

Thus, using the above-mentioned approximations we formally constructed, in the framework of the flat world, a $T$-invariant algebraic structure closed with respect to fields. However, the basic principle of the model is the assumption that a truly primary local world object is not a simple set of field multiplets, but their specially organized complex $U$ ($\{1\}, \{2\}$), including, together with the fields, the respective spin-group matrix bases in the fermion representation. Hence, the presumed fundamental-algebra equation should be closed with respect not only to the fields but also to the entire complex $U$. The calculation shows straightforwardly that for separate $L_u$- and $L_V$-algebras this is impossible, while in the $T$-invariant algebra based on the unified expression {12} such $U$-closedness appears to be possible under certain conditions. The most important of them is the restriction of the class of solutions of equation {12} abstracted from its derivation (with the results {11}) to the solutions with identified (so-to-say, at mutually spacelike points) pairs of complexes $U_1 \equiv U_{1'}$ and $U_2 \equiv U_{2'}$. Points $(1)$ and $(2)$ are connected with $L_u$-formalism fields and $(1')$ and $(2')$ - with $L_V$-formalism fields, and so we are speaking of identification in the framework of relations {12}, {11} of identical field components of different formalisms. The existence of such solutions in the flat Lagrangian world is obviously impossible. However, the equation thus occurring on two complexes and understood as a non-Lagrangian abstract algebraic relation can be put into correspondence with the geometrical images of the curved world. Indeed, a literally interpreted identification of fields in the conditional classical geometrical picture is reduced to identification of their arguments-points, i.e., to the gluing of mutually spacelike ends of the u- and V-generators intersecting at their interaction point and the closure of the corresponding light beams into a single "figure-of-eight" loop. Such 0-line deformation should naturally be regarded as resulting from the spontaneous occurrence, according to the above-described physical scheme, of an ultrashort-lived virtual black hole in the neighborhood of the field interaction act.

The GR equivalence principle might in this case mean that such jump-like transition from the flat to curved world, certainly understood as transition to the new model fields, at the same time leaves unchanged the coefficients of all the expressions derived in SR calculations of $L_u$- and $L_V$- algebras.

Another aspect of this procedure from the viewpoint of the Lagrangian scheme is due to the fact that equitype field components from different $L_u, L_V$-formalisms are dependent ("current") quantities in constructions in one formalism and independent ("field") quantities in the other. In this sense, when they are identified in the model beyond the framework of the Lagrangian formalism, a kind of "current-field" duality is realized. Along with algebra closure at two points, transition occurs from formally nonlinear commutator relations constructed from canonical fields on separate hyperplanes ($V=0$ and $u=0$) to equations for truly interacting – non-Lagrangian – fields. The causal relation in the thus constructed elementary link $((1),(2))$ corresponds to the $T$-jump between the $U_1$ and $U_2$ complexes (this jump connects the contributions of two opposite light-like motions to the bilinear operation) and is described by the fundamental algebra closed on it.

As a result of the calculations continuing (with allowance for field identification) the derivation of relation {11}, {12}, we obtain relations for the coefficients of the $U$ matrix providing $U$-closure of the algebra and establish its general structure.

The explicit form of this algebra, written already as an exact abstract algebra, is as follows ([17]):

$$\{U_1,U_2\} = S_1\left(\overline{U_1}^\tau + \overline{U_2}^\tau\right)S_2 \quad , \quad \{U_1,U_2\} \equiv -4ig\left([U_1,U_2]_- + [U_1^\tau,U_2^\tau]^{(\tau^{-1})}\right) \quad \{13\}$$

The $U$ matrix is defined in terms of the fields by expressions {1}, {2},

$$\overline{U} \equiv \Gamma^0 \overset{++}{U} \Gamma^0, \quad \Gamma^0 = \begin{pmatrix} 0 & \gamma^0 \times \delta_{ij} & 0 \\ \gamma^0 \times \delta_{ij} & 0 & 0 \\ 0 & 0 & g^{\mu\Rightarrow}_{\lambda\Downarrow} \end{pmatrix}, \text{ and } (\overline{U})^\tau \text{ is defined according}$$

to {10}.

We are led to $\overline{U}^\tau(\kappa_i) = U(\tilde{\kappa}_i)$, where $\widetilde{(\;)}$ is complex conjugation and $\kappa_i$, which are different roots of the equation $\kappa_i^3 = 1$, are factors of different fermion families. Owing to the property $\kappa_i^2 = \tilde{\kappa}_i$ both sides of equation {13} are consistent in these factors.

$$S_{1,2} = \begin{pmatrix} I \times \delta_{ij} & 0 & 0 \\ 0 & I \times \delta_{ij} & 0 \\ 0 & 0 & \alpha_{1,2} \cdot g^{\perp\varepsilon\Rightarrow}_{\perp\lambda\Downarrow} + \alpha'_{1,2} \cdot \left(n_{-,\lambda\Downarrow} n^{\varepsilon\Rightarrow}_+ + n_{+,\lambda\Downarrow} n^{\varepsilon\Rightarrow}_-\right) \end{pmatrix},$$

where $n_{\pm,\lambda}$ are light 4-vectors with properties {5} and $\alpha_{1,2}$, $\alpha'_{1,2}$ are unknown numerical coefficients to be found in the calculation of the correspondence principle.

The Lorentz-anisotropic matrix factors $S_1, S_2$ necessary for algebra closure are considered as carriers (within the limits of elementary link) of primary metric properties in the model, resembling $g_{\mu\nu}$ fields in GR. In the conditional geometrical picture these factors locally define the time axis $\frac{1}{\sqrt{2}}(n_{+,\mu} + n_{-,\mu})x^{\mu}$ and the invariant hyperplane $(n_{+,\mu} + n_{-,\mu})x^{\mu} = 0$ of local $T$-reflection that passes through the link symmetry center, i.e., they fix the local intrinsic frame of reference, existing in the given link, in which "instantaneous" equilibrium between the gravitational-inertial and gauge forces, expressed by algebra $U$-closure, is realized.

Closure of algebra {13} combines closure with respect to fields (taken together with their local bases), which is to provide an algebraic description of three-linear field interactions, and closure with respect to $U_1, U_2$ "points", which is put into correspondence with gravitational self-closure of the space-time neighborhood of a virtual local act of such an interaction.

## 6. Results

Thus, according to the given model, the constructed equation {13}, together with the associative realization of the noncommutative algebraic operation and the explicit expressions {1} and {2} for the $U$ matrix, determines the elementary causal proximity in the set of $U$-complexes of fields, and beyond the Lagrangian formalism it redefines the physical fields and forms a complete description of the submicroscopic structure of space-time considered as a unified virtual geometric-physical world.

One can in principle try to consider the obtained local algebra as a cell model in a quantum foam-like Wheeler-Hawking world [18], but with the essential difference that in the given scheme such a cell is always assumed to occur only in the neighborhood of a virtual physical-field interaction act.

According to the described procedure based on the correspondence principle with $L_u, L_V$ - Lagrangian formalisms of quantum field theory, the calculation of equation {13} also leads to a definition of numerical coefficients of the $U$-matrix, $S_1, S_2$-matrices, and group parameters in terms of two constants. In particular, we have

$$\frac{b_3}{b_1} = 3, \quad \frac{b_4}{b_1} = \frac{9}{2}.$$

Since the quantities $b_1, b_3, b_4$ introduced by formulas {3} are eigenvalues of the quadratic Casimir operator for irreducible representations of the charge group $\Omega$ ($b_1$ is such an eigenvalue for the adjoint representation), one can try to use group formulas (e.g., those from [19]) and to find from the obtained relations a simple compact group with the corresponding representations. Such a group appears to be group $E_6$ with adjoint representation of dimension $N_F = 78$ for vector fields, representation $[001010]$ $\left(\frac{b_3}{b_1} = 3\right)$ for the scalar field multiplet of dimension $N_\varphi = 26026$ and representation $[000200]$ $\left(\frac{b_4}{b_1} = \frac{9}{2}\right)$ for the fermion multiplet of dimension $N_\psi = 1337050$ \hfill {14}

(the numbers in square brackets are here coordinates of the highest weight of representation in the basis of fundamental weights).

Thus, the model predicts a concrete realization of the charge symmetry structure in the world of fields. True, when $E_6$ was proposed as a possible symmetry group [20], a 27-particle multiplet was assumed for fermions, and so the huge amounts of fermion and scalar degrees of freedom obtained in the model looks fairly exotic. But one should bear in mind that the manifestation of all the degrees of freedom of the system of fields is assumed to take place only at Planck energies. However, without calculations one cannot say how many of them will "survive" in the low-energy limit of the given model.

For the parameters $b_2, b_5$, the coefficients of the universal field matrix $U$, and matrices $S_1, S_2$ the following values are obtained:

$$b_2 = b_1,\ b_5 = 3b_1,\ a_{11} = a_{22} = -\frac{1}{2b_1},\ a_{13} = a_{23} = \widetilde{a}_{13},\ a_{31} = a_{32} = -\frac{1}{4b_1 a_{13}},\ c_{31} = c_{32} = -\frac{a_{31}}{2} = \frac{1}{8b_1 a_{13}},$$

$$y_1 = z_1 = -\frac{3}{5},\ y_2 = z_2 = \frac{19}{20b_1},\ \alpha_1 = \frac{21}{4},\ \alpha_1' = \frac{9}{4},\ \alpha_2 = \frac{3}{4},\ \alpha_2' = -\frac{9}{4} \qquad \{15\}$$

for undetermined $b_1 > 0$ and $a_{13} = \widetilde{a}_{13}$.

The yet unsolved group problem is possibly satisfaction of the equation $T^\varphi_{s,rr_1} q_r q^+_{r_1} = b_2 t_s$, which is critical for the model.

**7. Notes on a possible recurrent procedure generating a global causal structure. Continuous Lagrangian approximation at Planck energies and grand unification.**

Returning to equation {13} we note that when one of its arguments, for example, $U_1$ is given, the solution of the equation for the second argument may fail to be unique (perhaps even together with the values of the parameters $n_{\pm,\mu}$ from $S_1, S_2$, understood as eigenvalues of such a problem). Writing this ambiguity as $U_2^{(k)}$, $k = 1, 2, ..$, and restricting it to a (weak) analogue of the quantum-field locality condition

$$\{U_2^{(k)}, U_2^{(n)}\} = 0 \qquad \{16\}$$

we can consider the pairs $\left(U_1, U_2^{(k)}\right)$ as links of different causal paths commencing from one common $U_1$-point and already at an elementary step generating a set of new points (with a common reciprocity ratio {16}). In turn, taking each "newly born" point $U_2^{(k)}$ for the initial in the same consequent proximity equation, one can formally continue the construction of causal paths-chains branching (or possibly converging) at each next elementary stage, thus endowing the set of $U$-points with the structure of the discrete causal net commencing from one point (the Big Bang"?). The effective geometrical fields-connectednesses, including the macroscopic gravitational field, are then interpreted as the effect of dynamical degrees of freedom of the net itself. However, the detailing of these general considerations, in particular, deriving the equations of motion of such fields, requires knowledge of the method of solution to equation {13}, which is a problem in itself.

However, we can try to temporarily evade this problem and to use once again the assumption that the local unification of fields into a $U$-supermatrix is essential. Integrity of U means preservation of its physical meaning under any reversible covariance supertransformations [11] which preserve the fermion-boson grading of constituent cells of the $U$ matrix and the formal structure of equation {13}. Then we can suppose that in the correspondent continuous approximation the adequate Lagrangian must simply be expressed in terms of the integral U matrix (with the estimated coefficients!) and must be superinvariant. Such simplest expression appears to be $\sim str\left(\bar{U}^{(\tau^{-1})}U\right)$ and the corresponding density of the new Lagrangian $\check{L}$ is assumed to have the form $\check{L}=-Z\cdot str\left(\bar{U}^{(\tau^{-1})}U\right)$, $Z>0$, where $U$ is taken with the values of the coefficients {15} and with $E_6$-multiplets {14}. The gauge field expressions used in Sec. 2 in writing the Lagrangian density $L$ {4} are a priori taken as "momentum" variables $F_{\mu\nu,s}$, $\varphi_{\lambda,r}$, $\varphi_{\lambda,r}^+$, $\psi_\lambda$, $\overline{\psi_\lambda}$. As a result we have

$$\check{L}=\frac{Z}{b_1}\left[\frac{i}{2}\bar{\psi}\gamma^\lambda\psi_\lambda-\frac{i}{2}\overline{\psi_\lambda}\gamma^\lambda\psi+\bar{\psi}Cq_r\bar{\psi}\varphi_r-\psi C^{-1}q_r^+\psi\varphi_r^+-\frac{462825}{4}F_s^{\varepsilon\delta}F_{\varepsilon\delta,s}+\frac{462825}{1001}\varphi_r^\lambda\varphi_{\lambda,r}^+-\right.$$
$$\left.-b_1\frac{5779721574}{1001}\left(\varphi_r\varphi_r^+\right)^2-\frac{361}{100b_1}tr\left(q_{r_1}q_{r_2}^+q_{r_3}q_{r_4}^+\right)\varphi_{r_1}\varphi_{r_2}^+\varphi_{r_3}\varphi_{r_4}^+\right]. \quad \{17\}$$

This expression is considered as the effective Lagrangian density corresponding to the "gravitation-fluctuation" renormalization of the initial density L at the near-Planck energy level. It is precisely this expression that we are planning to use to calculate the degrees of freedom of a low-energy world through spontaneous symmetry breaking due to radiative corrections [21]. In this sense one may consider expression (17) as the effective starting-point for a certain model of grand unification.

**8. The possibility of an alternative view of the nature of time.**

We shall now regard the algebra-cell {13} from another viewpoint. From the construction one can see that it describes a kind of quantum jump between the nearest local past and future through an elusive "present", which (jump) is due to the virtual field interaction act. In the framework of the classical geometrical picture, under certain conditions this jump can visually be treated as the two-way $T$-reflection in time $U_1 \leftrightarrow U_2$ between the causally nearest $U$- points. The symmetrically two-way flowing time can on the whole be represented as a sequence of such local jumps. Accordingly, one can assume that it is not the (continuous) shifts but a discrete set of local acts of $T$-reflection that underlies all the $T$-reversible physical processes.

Thus, this model changes in a sense the idea of a possible nature of time.

A detailed presentation of the mathematical aspects of model construction can be found in [22].

9. Acknowledgements
I am grateful to V.O. Galkin and R.N. Faustov for their invariable friendly support.